\documentclass[twocolumn,dvipsnames]{aastex631}

\usepackage{apjfonts}
\usepackage{tabularx}
\usepackage{hyperref}
\usepackage{fancyhdr}
\usepackage{graphicx}
\usepackage{tabularx}
\usepackage{float}
\usepackage{amsmath}
\usepackage{refcount}

\usepackage{xcolor}

\graphicspath{{./}{figures/}}

\usepackage{apjfonts}

\DeclareSymbolFont{cmletters}{OML}{cmm}{m}{it}
\DeclareMathSymbol{v}{\mathalpha}{cmletters}{"76} 
\usepackage{hyperref}
\usepackage{booktabs}
\usepackage{amssymb}
\usepackage{amsmath}

\usepackage[sort&compress]{natbib}

\shorttitle{Self-consistent cylindrical jets from magnetized accretion onto black holes}
\shortauthors{Rohoza et al.}

\begin{document}

\title{How to Turn Jets into Cylinders near Supermassive Black Holes in 3D GRMHD Simulations}

\correspondingauthor{Valeriia Rohoza}
\email{valeriia.rohoza@u.northwestern.edu}

\author[0009-0000-4703-9808]{Valeriia Rohoza}
\affiliation{Center for Interdisciplinary Exploration \& Research in Astrophysics (CIERA), Physics \& Astronomy, Northwestern University, Evanston, IL 60202, USA}

\author[0000-0002-6883-6520]{Aretaios Lalakos}
\affiliation{Center for Interdisciplinary Exploration \& Research in Astrophysics (CIERA), Physics \& Astronomy, Northwestern University, Evanston, IL 60202, USA}

\author[0000-0003-3599-2161]{Max Paik}
\affiliation{Center for Interdisciplinary Exploration \& Research in Astrophysics (CIERA), Physics \& Astronomy, Northwestern University, Evanston, IL 60202, USA}
\affiliation{Courant Institute of Mathematical Sciences, New York University, New York, New York 10011, United States}
\author[0000-0002-2825-3590]{Koushik Chatterjee}
\affiliation{Black Hole Initiative at Harvard University, 20 Garden St., Cambridge, MA 02138, USA}
\affiliation{Center for Astrophysics, Harvard \& Smithsonian, 60 Garden St., Cambridge, MA 02138, USA}
\author[0000-0003-4475-9345]{Matthew Liska}
\affiliation{Center for Relativistic Astrophysics, Georgia Institute of Technology, Howey Physics Bldg, 837 State St NW, Atlanta, GA 30332, USA}
\affiliation{Institute for Theory and Computation, Harvard University, 60 Garden Street, Cambridge, MA 02138, USA}

\author[0000-0002-9182-2047]{Alexander Tchekhovskoy}
\affiliation{Center for Interdisciplinary Exploration \& Research in Astrophysics (CIERA), Physics \& Astronomy, Northwestern University, Evanston, IL 60202, USA}

\author[0000-0003-3115-2456]{Ore Gottlieb}
\affiliation{Center for Interdisciplinary Exploration \& Research in Astrophysics (CIERA), Physics \& Astronomy, Northwestern University, Evanston, IL 60202, USA}
\affiliation{Center for Computational Astrophysics, Flatiron Institute, New York, NY 10010, USA}
\affiliation{Department of Physics and Columbia Astrophysics Laboratory, Columbia University, Pupin Hall, New York, NY 10027, USA}

\begin{abstract}
Accreting supermassive black holes (SMBHs) produce highly magnetized relativistic jets that tend to collimate gradually as they propagate outward. However, recent radio interferometric observations of the 3C~84 galaxy reveal a stunning, cylindrical jet already at several hundred SMBH gravitational radii, $r\gtrsim350r_{\rm g}$. We explore how such extreme collimation emerges via a suite of 3D general-relativistic magnetohydrodynamic (GRMHD) simulations. We consider a SMBH surrounded by a magnetized torus immersed in a constant-density ambient medium that starts at the edge of the SMBH sphere of influence, chosen to be much larger than the SMBH gravitational radius, $r_{\text{B}}=10^3r_{\text{g}}$. We find that radiatively inefficient accretion flows (e.g., M87) produce winds that collimate the jets into parabolas near the BH. After the disk winds stop collimating the jets at $r\lesssim{}r_\text{B}$, they turn conical. Once outside $r_\text{B}$, the jets run into the ambient medium and form backflows that collimate the jets into cylinders some distance beyond $r_{\text{B}}$.  Interestingly, for radiatively-efficient accretion, as in 3C~84, the radiative cooling saps the energy out of the disk winds: at early times, they cannot efficiently collimate the jets, which skip the initial parabolic collimation stage, start out conical near the SMBH, and turn into cylinders already at $r\simeq300r_{\rm g}$, as observed in 3C~84. Over time, jet power remains approximately constant, whereas the mass accretion rate increases: the winds grow in strength and start to collimate the jets, which become quasi-parabolic near the base; the transition point to a nearly cylindrical jet profile moves outward while remaining inside $r_\text{B}$.
\end{abstract}


\section{Introduction}\label{section-introduction}
Highly energetic gas outflows from accreting supermassive black holes (SMBHs) span orders of magnitude in distance; for instance, relativistic collimated outflows, or jets, in M87 reach distances of $10$ kpc \citep{1999ApJ...520..621B}, whereas the X-ray jet in OJ287 extends out to $1$ Mpc \citep{2011ApJ...729...26M}. These outflows may impact the formation of galaxies by injecting energy and momentum into the ambient gas through the process known as the active galactic nucleus (AGN) feedback (see \citealt{2012ARA&A..50..455F}, \citealt{2017Natur.543...83P}). Understanding the AGN feedback is crucial for modeling galaxy evolution \citep{2018MNRAS.479.4056W}: jet activity can impact star formation \citep{2023MNRAS.518.4622E}, affect the cooling flows in galaxy clusters \citep{2023MNRAS.523.1104W}, and transport metal-enriched gas in galaxy cores \citep{2009ApJ...707L..69K}. Jets can produce high-energy (TeV) flares and accelerate hadronic cascades leading to multimessenger $-$ electromagnetic and neutrino $-$ emission, opening a novel window into the poorly understood dissipative processes around BHs (\citealt{1995APh.....3..295M}, \citealt{2012ApJ...749...63M}, \citealt{2018ApJ...864...84K}, \citealt{2018ApJ...863L..10A}, \citealt{2023ApJ...956....8F}). To meaningfully interpret these observations, it is important to understand how jets form and interact with the ambient medium.

Very long baseline interferometry (VLBI) observations have shed light on the properties of jet propagation, such as its acceleration and shape. \cite{2012ApJ...745L..28A} have found that the jet in M87 follows a parabolic collimation profile that transitions to conical at $r \simeq 10^5 r_{\text{g}}$, where $r_{\text{g}} = GM/c^2$ is the BH gravitational radius. This shape transition happens at the edge of the sphere of the gravitational influence of the SMBH, or the Bondi radius, defined as $r_{\text{B}} = GM/c^2_s$, where $c_s$ is the sound speed in the ambient medium. \cite{2012ApJ...745L..28A} and \cite{2017MNRAS.465.1608L} suggested that this transition from parabolic to conical in M87 might occur when the confining pressure profile changes. A study of a sample of 56 radio-loud AGNs shows the quasi-parabolic jet morphology inside the BH sphere of influence, which, due to the resolution constraints, may fit a parabolic-to-conical profile \citep{2017ApJ...834...65A}. \cite{2021A&A...647A..67B} find that the jet in NGC 315 exhibits a parabolic shape and becomes conical well inside the Bondi radius; they discuss that the winds launched by a thick accretion disk might be responsible for this shape transition. High-resolution images of 3C 273 show that the jet exhibits a parabolic shape up to $10^7r_{\text{g}}$ before expanding conically well outside of the BH sphere of gravitational influence \citep{2022ApJ...940...65O}, which they expect to be approximately of the same order of magnitude as $r_{\text{B}}$. The jet in Cygnus A features a parabolic shape up to $r \simeq 2.4 \times 10^4 r_{\text{g}} \simeq 0.05 r_{\text{B}}$, beyond which it turns into a cylinder; this suggests different environmental conditions than M87 \citep{2016A&A...585A..33B}. Although jets follow different collimation profiles at large radii, they typically have a parabolic shape near the BH. 

Surprisingly, recent observations of 3C 84 have revealed an almost cylindrical jet collimation profile at $350 r_{\text{g}} \lesssim r \lesssim 8,000 r_{\text{g}}$, where the jet maintains a near-constant cylindrical radius, $R_{\text{jet}} \approx 250 r_{\text{g}}$ \citep{2018NatAs...2..472G}. They conclude that a parabolic expansion starting at the base of the jet at the BH event horizon cannot produce such a wide jet so close to the BH. \cite{2023A&A...676A.114S} note that this is a recently restarted jet, which is about $10$ years old, and deduce that the cylindrical jet shape needs a nearly flat pressure profile of the ambient medium, which can be due to a ``leftover'' cocoon, the shocked ambient medium by the previous jet. They report the emission around the restarted jet on the (sub)parsec-scale from the putative cocoon, which can be responsible for the collimation. The cylindrical jet collimation profile so close to the BH challenges the commonly used models that typically feature the parabolic expansion of the jets near the BH.

Significant progress has been made in describing the jet formation and propagation analytically and numerically (e.g., \citealt{2005ApJ...620..878D}, \citealt{2006MNRAS.368.1561M}, \citealt{2007MNRAS.375..531M}, \citealt{2009ApJ...698.1570L}, \citealt{2011MNRAS.418L..79T}, \citealt{2019MNRAS.490.2200C}, \citealt{2023Galax..11...38C}; for reviews, see \citealt{2019ARA&A..57..467B}, \citealt{2020ARA&A..58..407D}). \cite{2011ApJ...740..100B} developed an analytical model of the relativistic unmagnetized jet propagating in the ambient medium. This model has been numerically verified in \cite{2018MNRAS.477.2128H} and recently generalized analytically and calibrated numerically to jet propagation in an expanding medium \citep{Gottlieb2022}. This model creates a framework for understanding the jet collimation profile: they characterize the jet shape based on the jet injection opening angle $\theta_0$ and the ratio of jet to rest-mass energy densities, $\tilde{L}$, at the location of the jet head. 

Yet, to understand the shape of the jet, it is crucial to account for the magnetic fields since they can modify jet collimation \citep{2009MNRAS.394.1182K} and give rise to 3D magnetic kink instabilities and dissipation (\citealt{1999MNRAS.308.1006L}, \citealt{2009ApJ...697.1681N}, 
\citealt{2016MNRAS.456.1739B}, \citealt{2023arXiv231011487L}). \cite{2016MNRAS.461L..46T} simulated relativistic magnetized jets launched by a spinning, perfectly conducting magnetized sphere. They immerse the sphere into a power-law density profile, mimicking the physical system outside the Bondi radius. However, their model did not include the central spinning BH that can launch jets magnetically via BH frame-dragging \citep{1977MNRAS.179..433B}. They also did not account for the accretion disk winds, which collimate the jets. \cite{2019MNRAS.490.2200C} carried out high-resolution 2D general relativistic magnetohydrodynamic (GRMHD) simulations with a large spatial separation ($5$ orders of magnitude) and measured the jet shape out to the Bondi radius scales. They found that the jet shape resembles the parabolic profile in M87 remarkably well throughout its entire length. Due to numerical constraints of 2D, \cite{2019MNRAS.490.2200C} adopts a SANE \citep[``standard and normal evolution'',][]{2012MNRAS.426.3241N}, state configuration,  where large-scale magnetic fields are sub-dominant, and jets are less variable \citep{2012MNRAS.426.3241N}. In contrast, a system where magnetic flux accumulates on the BH to the point it becomes dynamically important and obstructs accretion is known as magnetically arrested disk state (MAD; \citealt{2003PASJ...55L..69N}, \citealt{2011MNRAS.418L..79T}, \citealt{2022ApJ...941...30C}). Horizon-scale observations of M87 \citep{2019ApJ...875L...5E} are consistent with both SANE and MAD magnetic fields, but the polarization measurements favor MAD due to the presence of organized poloidal magnetic fields \citep{2021ApJ...910L..13E}. However, describing magnetic fields in AGN requires high-resolution polarimetric observations and inference from the numerical models, so the magnetic field remains a free parameter in the simulations.

A complete simulation of the full system in 3D, including jet launching by a rapidly spinning BH and propagation beyond the Bondi scales, is a numerically challenging task due to the large spatial and temporal separations (\citealt{2022ApJ...936L...5L}, \citealt{2023arXiv231011487L}). \cite{2021MNRAS.504.6076R} simulated the formation and propagation of jets in the systems with a non-rotating constant ambient medium, $r_{\text{B}} = 100 r_{\text{g}}$, and uniform magnetic fields tilted with respect to the BH spin axis. They found that the jets expand parabolically and then break apart due to the kink instability. \cite{2022ApJ...936L...5L} studied jet propagation out to $r_{\text{B}} = 10^3 r_{\text{g}}$, at the highest scale separation to date, and found that the jets, which started out parabolically near the BH, collimated into cylinders near the Bondi radius, i.e., at large distances from the BH. What causes jets to turn cylindrical near the BH remains a mystery. The nature of parabolic to conical shape transition at the Bondi radius also remains unclear: it has never been seen in GRMHD simulations that included the ambient medium with a well-defined $r_{\text{B}}$.

In this Letter, we investigate the dependence of the jet collimation profile on the ambient medium density and the thickness of the accretion disk via 3D GRMHD simulations. In Section \ref{section-simulation-parameters}, we describe our computational approach and simulation setup. In Section \ref{section-results}, we discuss how the jet shape depends on the properties of the accretion disk and ambient medium. We compare the results with the observations of 3C 84 and other AGN and discuss future work in Section \ref{section-discussion}. We conclude in Section \ref{section-summary}. We adopt units such that $G=M=c=1$, where $M$ is the mass of the BH; units of time then become, $r_{\text{g}} / c = 1$.

\section{Problem setup and computational approach}
\label{section-simulation-parameters}

We run the simulations using the GPU-accelerated GRMHD code H-AMR \citep{2022ApJS..263...26L}. We use spherical coordinates, $r$, $\theta$, and $\phi$, and a grid that is uniform in $\log{r}$, $\theta$, and $\phi$. We set the inner and outer boundaries of the computational grid at $r_{\text{in}}=0.8 r_{\text{H}}$ and $r_{\text{out}}=10^5 \, r_{\text{g}}$ respectively, where $r_{\text{H}} = \left( 1 + \sqrt{1 - a^2} \right) \, r_{\text{g}}$ is the event horizon radius, and $a$ is the dimensionless BH spin. Since we want to evolve the disk properly, which can be described as a non-relativistic gas, we adopt an ideal gas law equation of state with the polytropic index of $\gamma = 5/3$. Note that relativistic outflows are typically magnetically dominated and have low gas pressure, so the choice of the exact value for the polytropic index does not affect the outflow evolution significantly.

We start with a rapidly spinning BH of spin $a=
0.9375$ surrounded with an equilibrium hydrodynamic torus \citep{1976ApJ...207..962F}. We place the inner edge of the torus at $r_{\text{disk, in}} = 20 r_{\text{g}}$ and the pressure maximum at $r_{\text{disk, max}} = 41 r_{\text{g}}$. These parameters result in a torus with the outer edge just inside $r = 1,000 r_{\text{g}}$. We normalize the density such that $\text{max}\rho = 1$.
To initialize the magnetic field in the disk, we set the toroidal component of the magnetic vector potential to $A_{\phi} = \min \left( \rho - 0.2, \ 0 \right)$, which results in an initial poloidal magnetic field configuration. In this configuration, the magnetic flux surfaces follow the lines of constancy of density. We normalize the disk magnetic fields such that the ratio of the maximum thermal pressure to maximum magnetic pressure is $\max{p_{\text{gas}}} / \max{p_{\text{mag}}} = 100$.

Beyond the outer edge of the torus, we place a constant-density unmagnetized ambient medium: it starts at $r = 1,000 r_{\text{g}}$ and extends to the outer boundary of the grid. We set the internal energy of the gas such that the Bondi radius equals $r_{\text{B}} = 10^3 \, r_{\text{g}}$. 
We carry out a suite of three nonradiative simulations: M7, M6, and M5, which we name according to the ambient medium density, e.g., model M7 has the density $\rho_{\text{amb}} = 10^{-7}$ in the code units.
Since the luminosity, $L$, of 3C 84 is about $0.4 \%$ of its Eddington luminosity, $L_{\rm Edd},$ \citep{2014ApJ...797...66P}, it is close enough to the standard $1 \%$ Eddington ratio above which cooling becomes significant. In fact, already at such low luminosities, the disk becomes radiatively efficient (\citealt{2017ApJ...844L..24R,Liska2024,2023ApJ...944L..48L}). To account for this, we also carry out another simulation, M5C, where we introduce a prescription to approximate radiative cooling; here, ``C'' in the name stands for cooling. We decrease the internal energy of the disk over the local Keplerian timescale to a target temperature profile chosen to correspond to a target disk scale height \citep{2009ApJ...692..411N} that we choose to be $H / R = 0.1$.

All simulations have the resolution of the base grid, $(N_{r},\ N_{\theta}, \ N_{\phi}) = (288, \ 256, \ 192)$, and use $3$ levels of adaptive mesh refinement (AMR), which results in a maximum effective resolution of $(2304, \ 2048, \ 1536)$ in the jets. We use the AMR refinement criterion that activates depending on the number of the cells in $\theta$-direction in a jet or a cocoon (described in \citealt{2023arXiv231011487L}, \citealt{2022ApJ...933L...9G}); we identify a jet based on the maximum Lorentz factor (as defined in Section \ref{subsec:rjet}) and cocoon based on a proxy for entropy (see Appendix B in \citealt{2023arXiv231011487L}). We resolve the jet radius with at least $5$ cells out to $r = 10,000 r_{\text{g}}$.

\section{Results}\label{section-results}
\subsection{Jet collimation dependence on the ambient medium and disk cooling}
\begin{figure*}[!htp]
    \centering
    \includegraphics[width=0.75\textwidth]{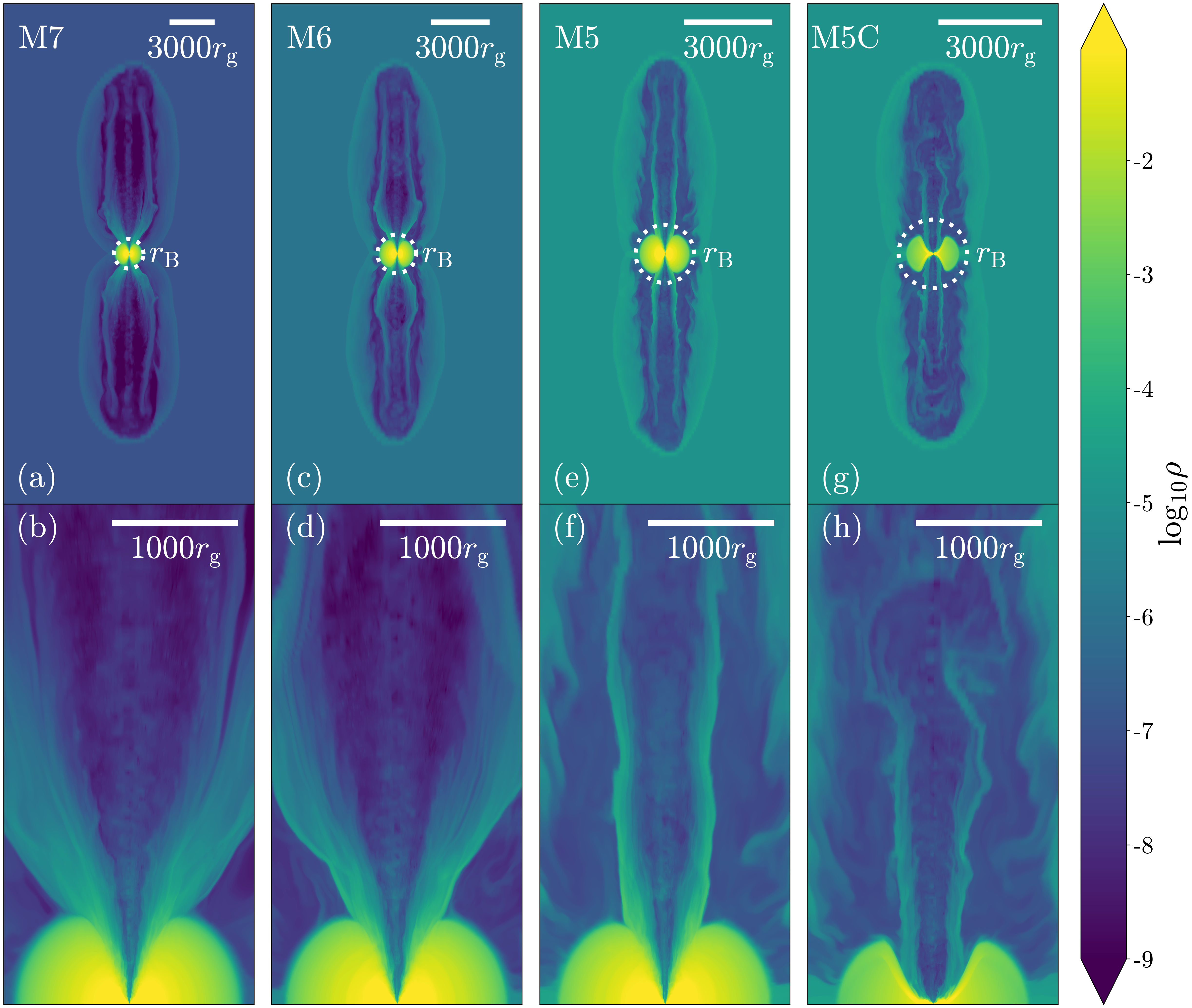}
    \vspace{5mm}
    \caption{Constant ambient medium collimates the jets which become cylindrical, with higher density resulting in a stronger collimation and narrower jet; surprisingly, the disk cooling allows jets to become cylindrical closer to the BH. Panels show 2D density slices in the $xz$-plane at $t=20 \text{k}  $ with each column corresponding to a different simulation. The top row is a zoom-out to show the full structure of the cocoon. The Bondi radius is shown with a dotted circle; note that each simulation has a different spatial scale. Panels (a), (c), and (e) show that as $\rho_{\text{amb}}$ increases, the jet propagates slower and becomes visually less stable. In panel (a), where $\rho_{\text{amb}}$ is lower, the cocoon is less prominent because the jet is stronger with respect to the ambient medium, and the jet creates a weaker shock. The jet interaction with the ambient medium redirects the material from the jet head and creates backflows, the low-density regions separated from the jet by more dense disk winds as seen in panels (a), (c), (e), and (g). The backflows are the most prominent in the highest-density simulations (e) and (g) since it is harder for the jet to drill through the ambient medium. The disk cooling causes the jet to become weaker and more prone to instabilities, and Panel (g) shows that the jet appears even less stable. The bottom row is a zoom-in between $z = 0 \ r_{\text{g}}$ and $z = 4,000 \ r_{\text{g}}$. Despite the difference in the large-scale jet morphology between models M7, M6, and M5 at $r \simeq 3,000 \ r_{\text{g}}$, panels (b), (d), and (f) show a similar jet morphology at the base, which is determined by the torus and its winds. With the cooling present, which impacts the torus, the jet in panel (h) opens up at smaller distances and appears wider. We notice that the backflows reach closer to the BH since the disk scale height decreases due to cooling.}
    \label{fig:density}
\end{figure*}

Figure \ref{fig:density} shows 2D density snapshots in the $xz$-plane in our simulations at $t = 20 \text{k}  $: each column corresponds to a different simulation; the top row shows the large-scale structure of the jets and their cocoons, while the bottom row shows the respective zoom in on the jet base. Figure \ref{fig:density}(a) shows that in our lowest ambient density model M7: by this time, the jet, which is launched by the BH, collides with and drills a hole in the constant density ambient medium. This interaction redirects the highly magnetized jet material at the jet head into backflows: these low-density (dark blue) regions are separated from the jet by the denser (light blue) disk winds. As the backflows return to the jet base, they collimate the jets into cylinders at $r \gtrsim 4,000 \ r_{\text{g}}$. As Figure \ref{fig:density}(b) shows, the jets transition to cylindrical geometry only well outside of the Bondi radius: they expand laterally closer to the BH, as we discuss below.
Figure \ref{fig:density}(c), (e) shows that models with higher ambient medium density, M6 and M5, exhibit more prominent backflows. The jets propagate slower as they struggle to drill through a denser ambient medium and become narrower due to a stronger collimation by a more energetic cocoon. In contrast, at distances within the Bondi sphere, which coincides with the characteristic size of the disk, the jets exhibit a shape similar to model M7, as we see in Figure \ref{fig:density} (d) and (f). This suggests that the disk winds determine the collimation profile of the jets near the BH, whereas the ambient medium impacts the jet farther away. Figure \ref{fig:density}(f) shows that even for the strongest backflows present in model M5, the disk and its winds prevent the backflows from reaching the jet base: the jet appears to turn cylindrical only after the disk ends. Thus, the fact that jet shape differences in models M7, M6, and M5 emerge only at large distances outside the torus indicates that the disk is a key factor in setting the collimation near the BH.

\begin{figure}[!htp]
    \centering
    \includegraphics[width=\columnwidth]{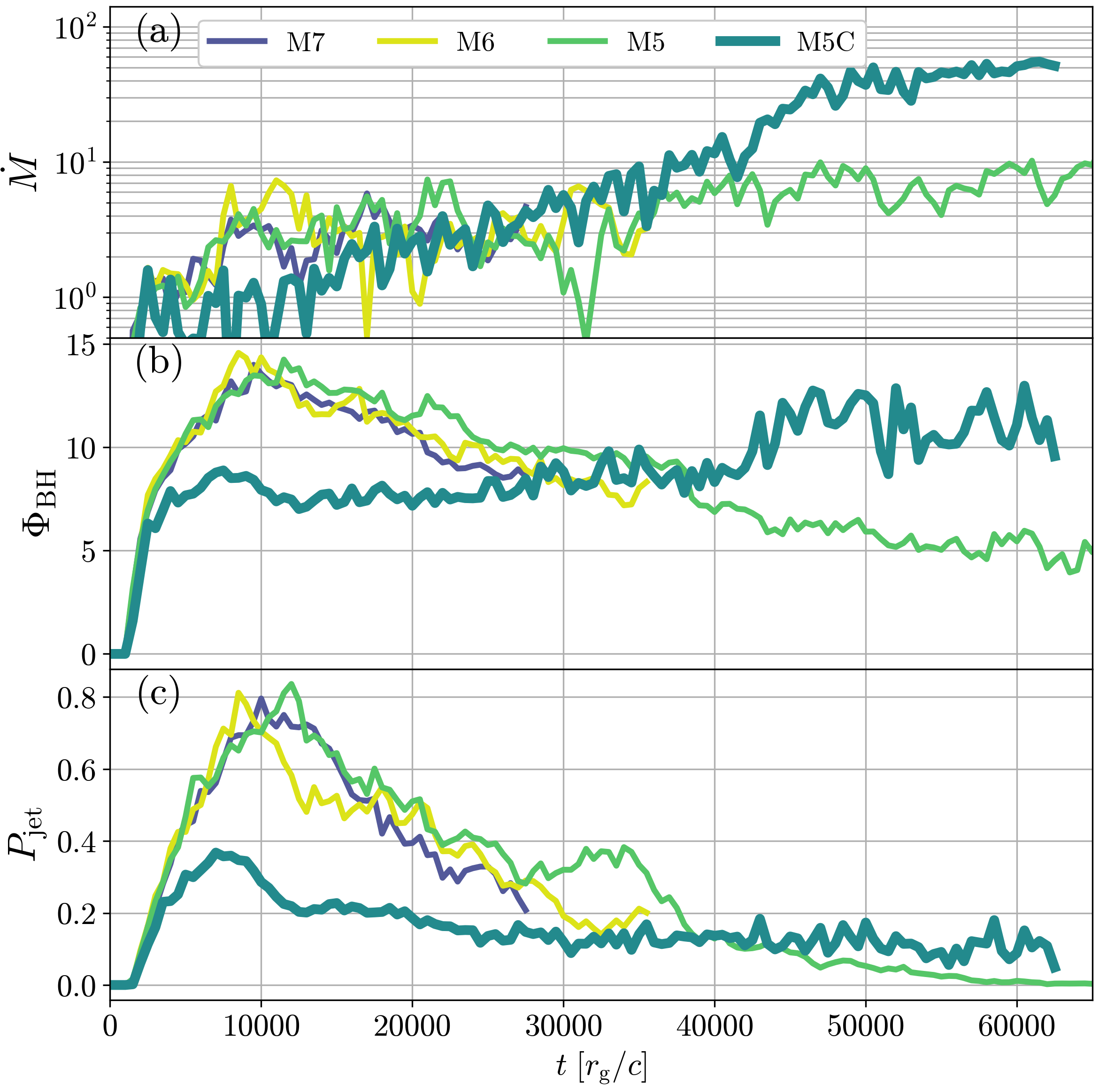}
    \caption{Whereas the ambient medium does not impact the power of the jet near its launching region, disk cooling makes a profound difference: it causes the mass accretion to increase continuously and the jets to become progressively weaker relative to the mass inflow. This allows us to study the dependence of the jet shape on the jet efficiency. All quantities are measured at $r=5 \ r_{\text{g}}$. Panel (a) shows that models M7, M6, and M5 reach a steady state by $t \simeq 15 \text{k}  $: the mass accretion rate $\dot{M}$ remains constant and approximately the same across the three simulations, M7, M6, and M5, because the torus and not the ambient medium dominates the BH accretion. In our model M5C, $\dot{M}$ flattens out at $t \simeq 15 \text{k}-30 \text{k}  $ and then rapidly increases at $ 30 \text{k} \lesssim t \lesssim 50 \text{k} $. Panel (b) shows similar smooth decline in $\Phi_{\text{BH}}$ across models M7, M6, and M5. In contrast, $\Phi_{\text{BH}}$ in model M5C remains constant at $10 \text{k} \lesssim t \lesssim 40 \text{k}$ before rising slightly at $t \simeq 40 \text{k}  $. Panel (c) shows the power of the jets, which is defined as the energy flux in strongly magnetized regions where the magnetization, $\sigma > 1$. The jet power is approximately the same for models M7, M6, and M5 and follows the expected trend, $P_{\text{jet}} \propto \Phi_{\text{BH}}$. As we see in Figures \ref{fig:m5c} and \ref{fig:rjetm5c}, at $t \gtrsim 35 \text{k}  $, the jet in model M5 starts to become dominated by the disk winds and disappears at $t \simeq 60 \text{k}  $. The power of the jet in model M5C starts flattening out at $t \simeq 15 \text{k}  $. The jet in model M5C is less powerful yet lives longer than the jet in model M5.\label{fig:pjet}}
\end{figure}
To understand how disk cooling impacts jet evolution, we consider model M5C with disk cooling. Figure \ref{fig:density}(g) shows that the jet in model M5C struggles to drill through the ambient medium, develops non-axisymmetric features, and creates even more prominent backflows than in model M5, suggesting that the jet becomes weaker and more prone to instabilities. Figure \ref{fig:density}(h) shows that the cooling causes the disk to collapse towards the mid-plane: this allows the backflows to reach smaller distances. The jet becomes wider, expands laterally at smaller distances, and develops a cylindrical shape closer to the BH, well inside the Bondi radius.

Since the jet behavior qualitatively depends on the strength of the jet with respect to the ambient medium, we quantify how strong the jet is. Measured at $r = 5 \ r_{\text{g}}$, Figure \ref{fig:pjet} shows the BH mass accretion rate, $\dot{M}$, the accumulated absolute magnetic flux on the BH, $\Phi_{\text{BH}}$, and the power of the jet, $P_{\text{jet}}$. We calculate the absolute magnetic flux on the BH as $\Phi_{\text{BH}} = 0.5 \iint |B_r| dA_{\theta \phi}$, where the integral is over the entire sphere with $r = 5 \ r_{\text{g}}$ and the factor of $0.5$ converts it to one hemisphere. To compute the power of the jet, we compute the energy outflow rate, $P_{\text{jet}} = \dot{M} c^2 - \dot{E}$, in the highly magnetized regions, where the magnetization exceeds unity, $\sigma = 2 p_{m} / \left(\rho c^2 \right) > 1$, where $p_m$ is the magnetic pressure and $\rho$ is the fluid frame mass density. All jet quantities are computed only for the top jet, so we constrain $\theta$, $0 \leq \theta \leq \pi / 2$; the bottom jet shows a qualitatively similar behavior.

Figure \ref{fig:pjet}(a) shows that models M7, M6, and M5 reach a steady state by $t \simeq 15 \text{k}  $ when the mass accretion rate reaches a quasi-steady state. With the mass accretion rate staying approximately constant and the magnetic flux decreasing, as seen in Figure \ref{fig:pjet}(b), the power of the jets in Figure \ref{fig:pjet}(c) decreases as expected \citep{2011MNRAS.418L..79T}. Since models M7, M6, and M5 exhibit very similar behavior in $\dot{M}$, $\Phi_{\text{BH}}$, and $P_{\text{jet}}$, we conclude that the differences in the ambient medium density do not affect the jet dynamics, which instead is entirely determined by the accretion disk. Over time, as we see in Figure \ref{fig:pjet}(c), the jet in our highest ambient density model M5 becomes progressively weaker until it disappears completely at $t \simeq 60 \text{k}  $.

\begin{figure*}[!htp]
    \centering
    \includegraphics[width=0.75\textwidth]{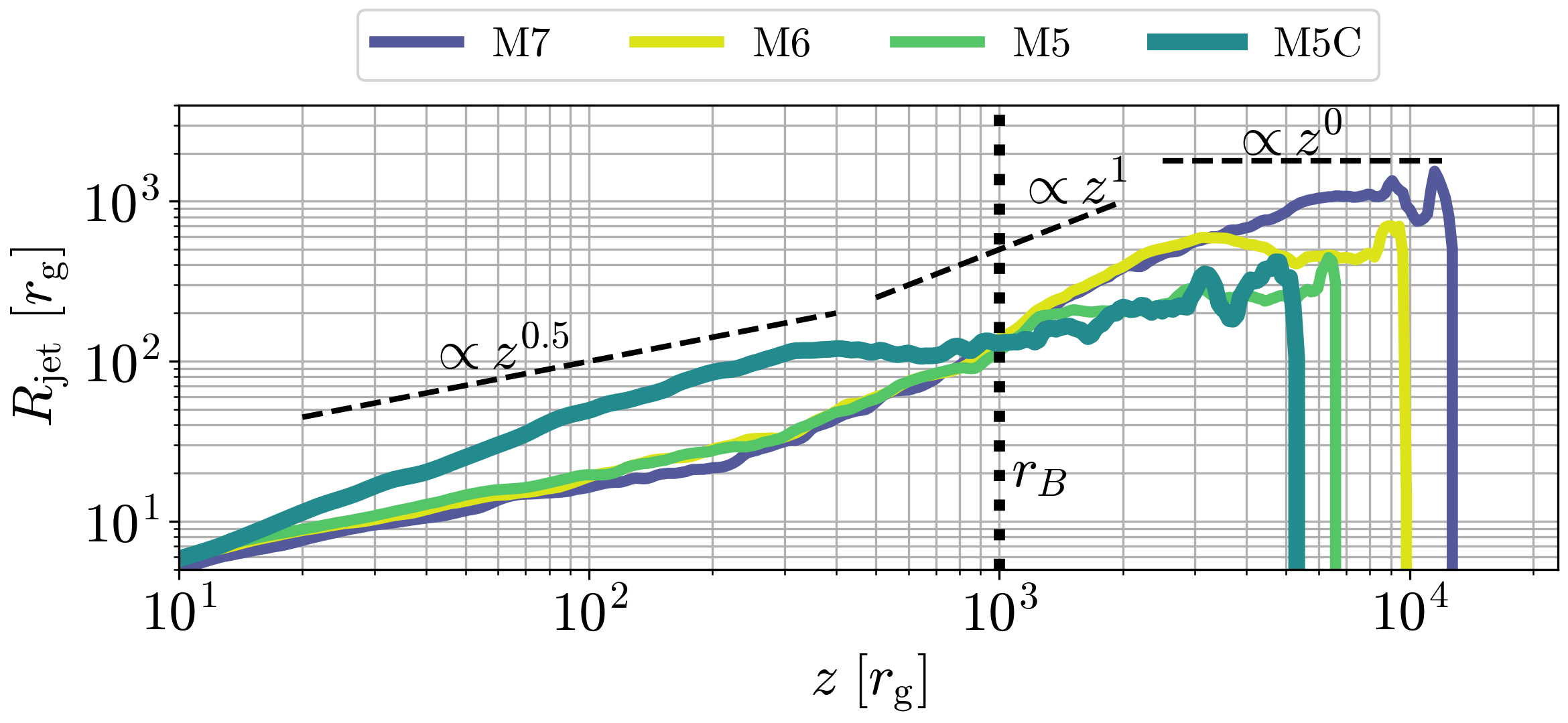}
    \caption{First self-consistent demonstration of the jet shape transition from parabolic to conical to cylindrical in models M7, M6, and M5, GRMHD simulations, at the largest to date Bondi-to-gravitational radius scale separation of $3$ orders of magnitude. The disk cooling in model M5C causes the jet to start out conical and transition to a cylindrical shape with $R_{\text{jet}} \simeq \left( 100-200 \right) \ r_{\text{g}}$ well inside the Bondi radius, already at a distance of $r \simeq 300 \ r_{\text{g}}$. The M5C jet radius approximately matches the observed values for the 3C 84 jet, where $R_{\text{jet}}^{\text{3C 84}} \gtrsim 250 \ r_{\text{g}}$ at $r^{\text{3C 84}} \gtrsim 350 \ r_{\text{g}}$ \citep{2018NatAs...2..472G}. This figure shows the collimation profile of the jet as a function of the distance along the jet at $t=20\text{k}$, when measured $\dot{M}$ are similar and $P_{\text{jet}}$ are within a factor of $2$. Jets in models M7, M6, and M5 have a similar parabolic collimation profile extending up to $r \simeq 200 \ r_{\text{g}}$, beyond which they expand conically. Outside of the sphere of the BH gravitational influence, the jets in models M7, M6, and M5 become cylindrical: as we increase the ambient medium density, the transition point moves inward. The jet in model M5C initially expands conically up to around $300 \ r_{\text{g}}$ and then becomes nearly flat.\label{fig:rjet}}
\end{figure*}
Unlike the changes in the ambient medium density, disk cooling significantly impacts the near-BH behavior of the system: Figure \ref{fig:pjet}(a) shows that the mass accretion rate in model M5C flattens out at $t \simeq 15 \text{k}-30 \text{k}  $ at the level similar to models M7, M6, and M5; afterward it starts increasing until it saturates at $t \simeq 50 \text{k}  $. We will use the similarity in mass accretion rate across all models and compare the jet collimation profiles at $t = 20 \text{k}  $ (Section \ref{subsec:rjet}).

As we will see in Section \ref{subsec:time-evol-m5c}, at $t \gtrsim 30 \text{k} $, more of the disk in model M5C loses thermal support, collapses towards the mid-plane, and moves closer to the BH. Although this results in a significant increase in the mass accretion rate, Figure \ref{fig:pjet}(c) shows that the power of the jet approximately flattens out at $t \gtrsim 15 \text{k}  $. This leads to jet efficiency, $\eta_{\text{jet}} = P_{\text{jet}} / \left( \dot{M} c^2 \right)$, decreasing in time. This allows us to study the jet shape evolution for a range of $\eta_{\text{jet}}$ values (Section \ref{subsec:time-evol-m5c}).

\subsection{Jet radius comparison}
\label{subsec:rjet}

To compute the jet radial collimation profile, we define a jet using the parameter $\mu = - T^r_t / \left(\rho u^r \right)$, where $\left( - T^r_t \right)$ is the total energy flux in the radial direction, where $T^{\mu}_{\nu}$ is the stress-energy tensor, $\rho u^r$ is the radial rest mass flux density, and $u^r$ is the contravariant radial component of the proper velocity 4-vector. The $\mu$ parameter can be thought of as the maximum possible Lorentz factor the jet could accelerate to if all of its energy flux were to convert into kinetic energy flux. To avoid division by zero where $u^r$ vanishes, we use an alternative form of $\mu$ given by the total specific energy of a fluid (\citealt{2019MNRAS.490.2200C}, \citealt{2022ApJ...936L...5L}): $\mu = -u_{t} (h + \sigma + 1)$, where $u_t$ is the temporal covariant 4-velocity component, which is a conserved quantity for point masses, $h=\left(u_{\text{g}} + p_{\text{g}}\right) / \left(\rho c^2 \right)$ is the specific (non-relativistic) gas enthalpy, which excludes the rest-mass contribution, and magnetization is approximated as $\sigma = 2 p_{m} / \left( \rho c^2 \right)$ (see \citealt{2023arXiv231011487L}). We define the jet as $\mu \geq 5$ for model M6 and $\mu \geq 4$ for models M7, M5, and M5C for Figure \ref{fig:rjet}. We determine the $\mu$ cutoff by examining its angular distribution: below these values, it sharply drops to $\sim 1$, which corresponds to non-relativistic gas. To exclude backflows, we consider regions with positive radial velocity directed away from the BH.

Figure \ref{fig:rjet} shows the jet radius, $R_{\text{jet}}$, as a function of distance from the jet base, $z$. We calculate the jet radius as
\begin{equation}
    R_{\text{jet}} = r \sin{\theta_{\text{jet}}}
\end{equation}
where $\theta_{\text{jet}}$, is the jet opening angle, calculated from the area subtended by the jet: 
\begin{equation}
    \theta_{\text{jet}} = \arccos{\left(1 - \frac{1}{2 \pi r^2}  \ \iint dA_{\phi \theta \text{, jet}}\right)}
\end{equation}

Figure \ref{fig:rjet} depicts the jet radius as a function of distance from the BH. For model M7, the jet expands parabolically up to $200 \ r_{\text{g}}$; this profile is characteristic of jets launched via the Blandford-Znajek mechanism due to the collimation by the disk winds (\citealt{1977MNRAS.179..433B}; see also \citealt{2007MNRAS.375..531M}, \citealt{2019MNRAS.490.2200C}). When the jet outruns the disk winds, they cannot keep collimating the jet farther, and it expands laterally and transitions into a conical shape. However, such a free, ballistic expansion cannot last forever in the presence of the ambient medium. Eventually, at $r \simeq 5,000 \ r_{\text{g}}$, due to the constant pressure environment formed by the backflows, the jet gets collimated into a cylinder, as predicted in the asymptotic solution by \cite{2009ApJ...698.1570L}. Here, we demonstrate, for the first time, that such a shape transition naturally emerges in jets self-consistently launched by a rotating accreting BH.

Similar to model M7, the jets in models M6 and M5 exhibit a parabolic shape near the jet base, inside the Bondi radius. Figure \ref{fig:rjet} shows that their shape also transitions to the conical shape at $r \simeq 200 \ r_{\text{g}}$. However, Figure \ref{fig:rjet} also reveals that the jets in models M6 and M5 turn into cylinders closer to the BH, at $r \simeq 3,000 \ r_{\text{g}}$ and $r \simeq 1,500 \ r_{\text{g}}$, respectively: denser ambient medium results in stronger backflows that can collimate the jets closer to the BH. In addition to collimating jets into cylinders closer to the BH, models M6 and M5 feature narrower cylindrical jets because the jets become less powerful relative to the denser ambient medium. 

Interestingly, Figure \ref{fig:rjet} shows that the jet in model M5C expands right from the outset and becomes significantly wider than in model M5 on the scales of a few hundred $r_{\text{g}}$, which is the characteristic size of the torus. We associate such rapid jet expansion at small radii with the disk cooling that weakens the disk outflows and suppresses the jet collimation. However, already at $r \simeq 300 \ r_{\text{g}}$, the jet in model M5 becomes cylindrical due to the collimation by the backflows. In fact, in model M5C, the backflows manage to get closer to the jet base and turn the jet into a cylinder on a smaller scale than in other models due to the cooling-reduced scale height of the disk and weakened disk outflows. The transition into a cylindrical shape happens well inside the Bondi radius and may relate to the observations of the cylindrical jet in 3C 84 at similar distances, $r \gtrsim 350 \ r_{\text{g}}$ \citep{2018NatAs...2..472G}. The cylindrical radius of our jet in model M5C, $R_{\text{jet}}^{\text{sim}} \simeq (100-200) \ r_{\text{g}}$, is within a factor of $\simeq 2$ of the observed value, $R_{\text{jet}}^{\text{3C 84}} \gtrsim 250 \ r_{\text{g}}$.

\subsection{Time evolution of radiatively efficient model, M5C} \label{subsec:time-evol-m5c}

\begin{figure}
    \centering
    \includegraphics[width=\columnwidth]{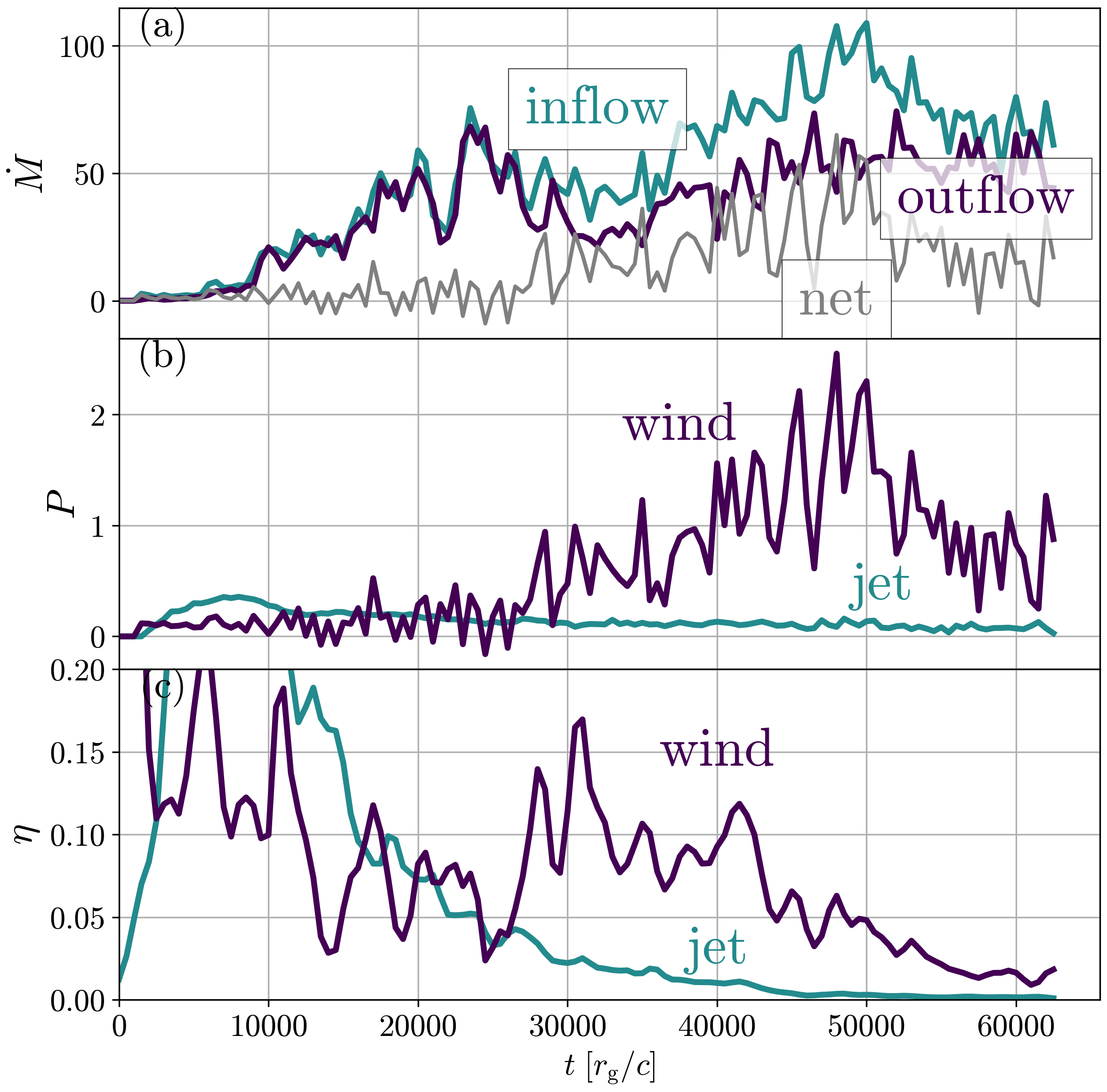}
    \caption{The disk cooling in model M5C allows the jets to energetically dominate the disk wind initially, but over time, the mass accretion rate increases, and the winds become stronger, whereas the jet power slowly decreases, because it is limited by the amount of available large-scale vertical magnetic flux in the system. All quantities are measured at $r=15 \ r_{\text{g}}$ to capture the disk winds and the jet but exclude backflows. Panel (a) shows the inflow, outflow and net mass accretion rates. Until $t \simeq 30 \text{k}  $, the mass inflow and outflow rates are comparable. At $t \gtrsim 30 \text{k}  $, due to cooling, the disk material comes closer to the BH and its accretion timescale decreases. More material starts falling in while approximately the same amount of material flies out, which results in an increased inward net mass flux. In panel (b), at $t \simeq 30 \text{k}  $, the power of the wind becomes larger than the power of the jet, and the winds become more prominent. The power of the jet reaches an approximately constant value at $t \simeq 15 \text{k}  $ and slowly decreases over time. Panel (c) shows the jet and wind efficiencies $\eta_{\rm j}$ and $\eta_{\rm w}$. The efficiency of the jet drops drastically due to the increasing mass accretion rate, which allows us to analyze the jet shape at different jet efficiencies.} \label{fig:m5c}
\end{figure}
Because the disk winds are crucial for jet collimation inside the Bondi radius, and disk cooling impacts the disk structure, we study how the cooling impacts the mass flow rates and powers of the outflows. Figure \ref{fig:m5c}(a) shows the magnitudes of mass inflow and outflow rates, $\dot{M}_{\text{in}}$ and $\dot{M}_{\text{out}}$. We define inflow and outflow based on the sign of radial velocity, $u^{r}$: inflow has negative, and outflow has positive radial velocities, respectively. We adopt the sign convention that the mass accretion rate is positive if directed into the BH and negative if it is directed away from the BH. Based on this definition, the net mass accretion rate equals to $\dot{M}_{\text{net}} = \dot{M}_{\text{in}} - \dot{M}_{\text{out}}$. Figure \ref{fig:m5c}(a) shows that at $t \lesssim 30 \text{k}$, the mass inflow follows the mass outflow, resulting in a relatively low net mass accretion rate, as seen in Figure \ref{fig:pjet}(a) at $t \lesssim 30 \text{k}$. At $t \simeq 30 \text{k}  $, the mass inflow rate starts increasing relative to the mass outflow rate. We can interpret this increase in $\dot{M}_{\text{in}}$ as follows: as the disk loses thermal pressure support due to radiative cooling and collapses, more mass comes closer to the BH, and the mass accretion timescale decreases: $\dot{M}_{\text{in}}$ and $\dot{M}$ both increase, as seen in Figs.~\ref{fig:pjet}(a) and \ref{fig:m5c}(a). At $t \simeq 30 \text{k}  $, Figure \ref{fig:m5c}(b) shows that the disk winds become more powerful and start dominating the jets. As seen in Figure \ref{fig:m5c}(c), the jet energy efficiency decreases as expected due to a nearly flat jet power and an increasing mass accretion rate.

\begin{figure*}
    \centering
    \includegraphics[width=0.75\textwidth]{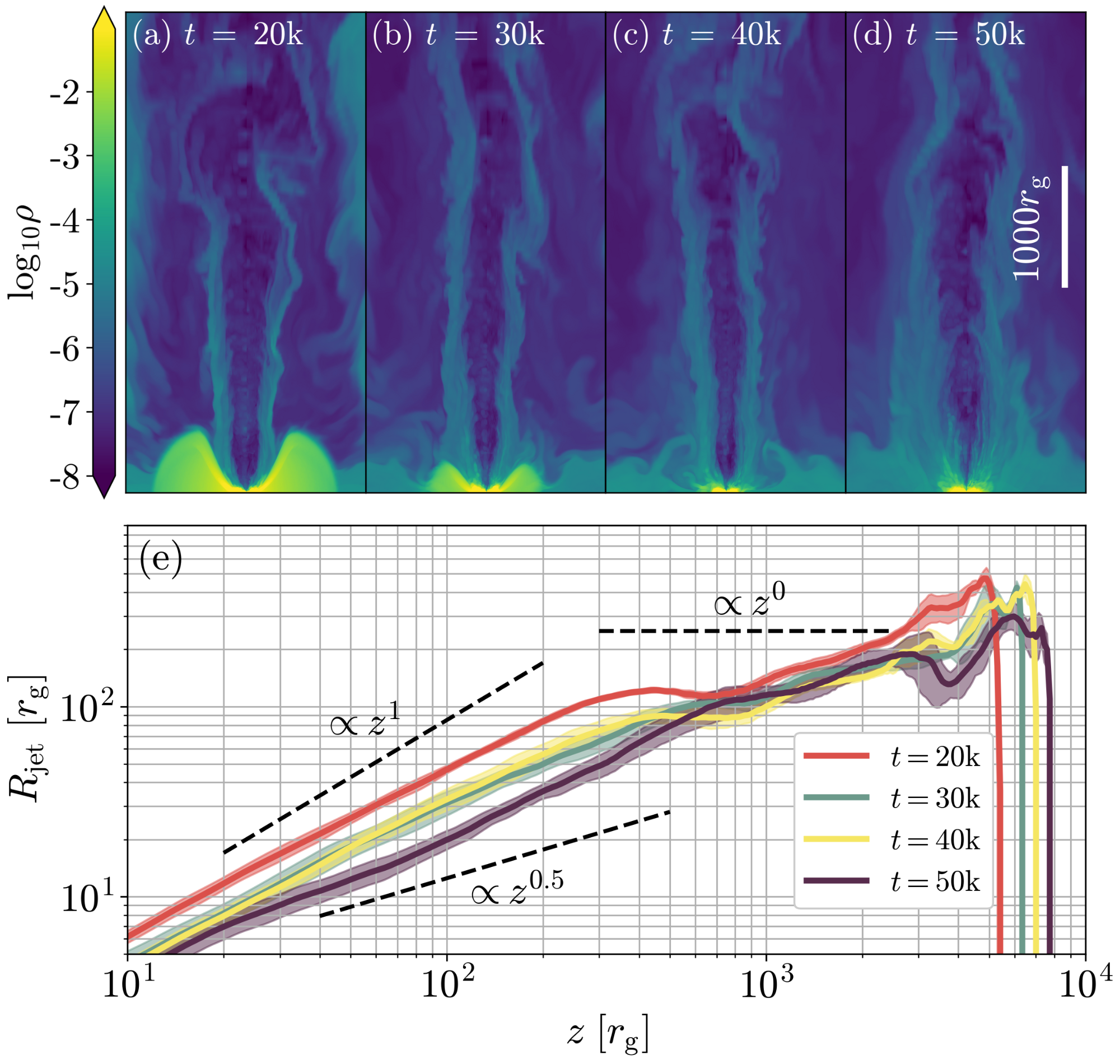}
    \caption{As the winds grow stronger, the jet in model M5C transitions from quasi-conical to quasi-parabolic shape but remains cylindrical starting from $r \simeq 700 \ r_{\text{g}}$. Panels (a)-(d) show density slices in the $xz$-plane at $t = 20 \text{k}   - 50 \text{k}  $ every $\Delta t = 10 \text{k}  $. The disk becomes visibly smaller and opens up the space for the jet expansion since it collapses towards the mid-plane due to cooling. It also allows the backflows to come closer to the jet base and turn the jet into a cylinder. However, panels (a)-(d) show that the wind launched from the disk becomes more prominent, opposes the backflows, and strongly collimates the jet. Panel (e) shows the jet radial profile, with each line corresponding to a different time, where we define the jet based on $\mu \geq 3$. The solid line is a value of the radius of the jet at time $t$ time-averaged over $2 \ \text{k}  $; the shaded region shows the values within one standard deviation from the average. At $t = 20 \text{k}  $, the jet exhibits a wide cylindrical jet starting from $r \simeq 200 \ r_{\text{g}}$. However, as the power of the jet remains approximately constant and the power of the wind increases, the jet becomes quasi-conical at $t = 40 \text{k}  $ and $t = 50 \text{k}  $. At $t = 50 \text{k}  $, it gets collimated into a quasi-parabolic shape up to $r \simeq 700 \ r_{\text{g}}$ and becomes nearly flat. Note that the jet becomes unstable at the time $t \geq 60 \text{k}  $.\label{fig:rjetm5c}}
\end{figure*}
To understand how the jet shape changes as a function of jet and wind energy efficiencies, in Figure \ref{fig:rjetm5c}, we analyze the time series of model M5C snapshots, taken from $t=20 \text{k}  $ to $t=50 \text{k}  $, after which the jet disappears, suppressed by the winds and the ambient medium. Figure \ref{fig:rjetm5c}(a) shows that the outer edge of the torus reaches $r \simeq 500 \ r_{\text{g}}$, and only the inner regions of the torus are significantly impacted by the cooling. Because the inner disk regions collapse towards the mid-plane and the winds weaken due to radiative cooling, the jet expands nearly conically near the BH, as seen in Figure \ref{fig:rjet}. At later times, cooling starts to impact the outer parts of the disk as well, and a sequence of panels (b), (c), and (d) in Figure~\ref{fig:rjetm5c} shows how the disk reduces in size over time. 

Figure \ref{fig:rjetm5c}(e) shows jet collimation profiles time-averaged over the time interval of $\Delta t = 2 \text{k}$ centered around the times shown (see legend); the shaded regions show one standard deviation. In Figure~\ref{fig:rjetm5c}(e), while computing the collimation profile of M5C, we use $\mu \geq 3$ for the jet definition since it becomes progressively weaker and, therefore, its values of $\mu$ decrease. Figure \ref{fig:m5c} shows that the winds become stronger than the jet at $t \gtrsim 30 \text{k}$. Figure \ref{fig:rjetm5c}(e) shows that as a result, the winds more strongly collimate the jets at $30 \text{k} \lesssim t \lesssim 50 \text{k}$ at $r \lesssim 500 \ r_{\text{g}}$ than at $t = 20 \text{k}$. 
Note that the rise of the wind power at $t \simeq 30 \text{k}$ pushes the backflows away from the jet base, and causes the jet to become quasi-parabolic near the BH and turn into a cylinder farther out. The jet profile at later times shows increasingly more variability since it becomes less stable. Despite the stronger collimation near the BH, the jet still ends up turning cylindrical inside $r_{\rm B}$, albeit at larger distances, $r \simeq 400 - 700 \ r_{\text{g}}$.

\section{Discussion}\label{section-discussion}
\subsection{Comparison with observations}
Because the density unit in our simulation is free, to map our scale invariant simulations to real systems, we calculate the jet stability parameter in physical and code units as derived in \cite{2016MNRAS.456.1739B} and \cite{2016MNRAS.461L..46T},
\begin{equation}
    \Lambda = K \times \left( \frac{P_{\text{jet}}}{n m_{p} r^2 \gamma_{\text{jet}}^2 c^3} \right)^{1/6} \label{eq:param}
\end{equation}
where $P_{\text{jet}}$ is the jet power, $n$ is the ambient medium number density, $m_{p}$ is the proton mass, $r$ is the distance to the jet head, $\gamma_{\text{jet}}$ is the Lorentz factor of the jet fluid, and $K$ is a constant factor. Qualitatively, eq. \eqref{eq:param} involves the power of the jet in the numerator and the density of the ambient medium times the length scale squared in the denominator. We consider the length scale equal to the Bondi radius since it is the order of magnitude of the jet length in the system. Therefore, we can compute a dimensionless ratio of the quantities that determines the jet stability:
\begin{equation}
    \lambda = \frac{P_{\text{jet}}}{n m_{\text{p}} r_{\text{B}}^2 c^3}
    \label{eq:lambda}
\end{equation}
By computing the simulated and observed values of $\lambda$, we can make quantitative comparisons of jet power normalized by their ambient density between simulations and observations.
In the code units, 
\begin{equation}
\lambda_{\rm sim}= 0.05
        \times \left(\frac{P_{\rm jet}}{0.5}\right)
        \times\left(\frac{\rho_{\rm amb}}{10^{-5}}\right)^{-1}
        \times\left(\frac{r_{\rm B}}{10^3}\right)^{-2}.
\label{eq:lambda_sim}
\end{equation}
For 3C~84, we use the mass of the BH, $M = 8 \times 10^8 \ M_\odot$ \citep{2013MNRAS.429.2315S}, the ambient medium number density, $n = 0.05 \ \text{cm}^{-3}$, the Bondi radius, $r_{\text{B}} =  8.6 \ \text{pc}$, and the power of one jet, $P_{\text{jet}} = 2.8 \times 10^{43} {\rm erg\,s^{-1}}$ (model FL in \citealt{2016MNRAS.455.2289F}) to compute the stability parameter, $\lambda_{\rm 3C84}\sim0.02$, using eq.~\eqref{eq:lambda}. Because for 3C~84, $L/L_{\rm Edd} \simeq 0.4 \%$ \citep{2014ApJ...797...66P}, which is sufficiently close to the canonical $1\%$ divide between radiatively-inefficient and radiatively-efficient flows, we expect the accretion flow in 3C~84 to be to some degree radiatively efficient. Hence, our model M5C provides the most relevant comparison. Indeed, $\lambda_{\rm 3C84}$ is very close to that in model M5C, $\lambda_{\rm M5C}=0.02$ (the stability parameters for jets in M6 and M7 are $10$ and $100$ times higher, respectively). 

Thus, our radiatively efficient model M5C offers insight into what can turn 3C~84 jets into cylinders so close to the BH. If the disk outflows in 3C~84 are weakened by the radiative cooling of the accretion disk, as in our model M5C, they will provide weak collimation to the jet and let it expand ballistically, i.e., essentially radially (Figure ~\ref{fig:rjetm5c}). The weakened and weakly collimated disk winds are unable to put up a fight against the powerful backflows, either: produced by the jet interaction with the ambient medium, the backflows push the winds out of the way and reach deep inside the Bondi radius, where they collimate the jets into cylinders. The significant impact of the backflows on the jet shape deep inside $r_{\text{B}}$ in our model M5C suggests that they may be important to the jet collimation in real systems well inside the Bondi radius. In future work, we will compute synthetic images of jets and backflows to determine their observational signatures and compare them to the images of the 3C~84 jet and cocoon.

Interestingly, whereas our jets collimate into cylinders at about the same distance as in 3C~84, they are about twice as narrow, $R_{\rm jet}^{\rm sim}\simeq (100-200)r_{\rm g}$, as those in 3C~84, $R_{\rm jet}^{\rm 3C84} \gtrsim 350r_{\rm g}$ \citep{2018NatAs...2..472G}. 
\citet{2023A&A...676A.114S} note that the jet in 3C~84 has recently restarted and is only 10 years old. This duration translates into $t = 76{\rm k}\, t_{\rm g}$, which is comparable to the duration of our model M5C and is about 4 times as long as $t=20$k, at which we measured the jet width (Sec.~\ref{subsec:rjet}). If the mass accretion rate in our model M5C did not increase in time, we would expect that by $t=76$k, our simulated jets would gradually expand laterally, and their width would more closely resemble that observed in 3C~84.

For M87, we adopt the mass of the BH, $M = 6.5 \times 10^9 \ M_\odot$ \citep{2019ApJ...875L...1E}, the number density at $r=1$~kpc, $n \simeq 0.2\,\text{cm}^{-3}$, and the Bondi radius, $r_{\text{B}} =  (0.11\pm0.02) \, \text{kpc}$ \citep{2015MNRAS.451..588R},\footnote{Note that their definition of the Bondi radius is twice as large as ours.\label{ftn:rb_def}} and the jet power, $P_{\text{jet}} \sim 10^{44} \, {\rm erg\, s^{-1}}$\citep{1996ApJ...467..597B,2000ApJ...543..611O}.  Plugging these values into eq.~\eqref{eq:lambda} gives $\lambda_{\rm M87} =10^{-4}$.  The accretion flow in M87 is highly sub-Eddington, $L/L_{\rm Edd}\sim 10^{-6}$ \citep[e.g.,][]{2015ApJ...809...97B} and thus has a low radiative efficiency. Thus, we perform the comparison against our non-radiative models, M7--M5. We find that the jet stability parameter in model M5 at $t \simeq 20 \text{k}$ is $\lambda\simeq 0.05$, i.e., a factor $\simeq 500$ larger than in the M87 jet. Thus, our jets are much more powerful than in M87. Despite this, we can try to draw conclusions from our models for the case of M87. In our models M7, M6, and M5, the jets consistently develop a shape transition from parabolic to conical before becoming cylindrical around the Bondi radius. This suggests that in M87, the disk winds would collimate the jets out to a fraction of the Bondi radius, after which the jets would first turn conical and, shortly outside $r_{\rm B}$, cylindrical. However, the transition to cylindrical is not observed in M87. It is possible that for weaker jets, 3D magnetic kink instabilities can modify the picture found in this work: at smaller jet powers, the jets will be less stable, are more likely to break apart due to the magnetic kink instability, and are unable to propagate beyond a critical distance \citep{2016MNRAS.461L..46T}. Once the jets become stuck, their exhaust outruns the jet head; as a result, the backflows do not form and cannot collimate the jets into cylinders. Whether this leads to a conical shape beyond the Bondi radius, as observed in M87, will be the topic of future work. Note that because $P_{\rm jet}$ in model M5 decreases with increasing time such that by $t=60$k the stability parameter drops down to $\lambda_{\rm M5,60k}\lesssim10^{-3}$, late-time evolution of model M5 might provide a better quantitative comparison to M87. We leave this to future work.

For NGC 315, the inferred BH mass ranges from $M=3\times 10^8M_\odot$ \citep{2007ApJ...671L.105G} to $M=3.4\times 10^9M_\odot$ \citep{2003A&A...399..869B}. Following \citet{2021A&A...647A..67B}, we adopt an intermediate BH mass, $M = 1.3 \times 10^9 \, M_\odot$ \citep{2005ApJ...633...86S}. We take the Bondi radius to be $r_{\text{B}} = (51\pm 25) \ \text{pc}$ \citep{2020ApJ...894..141I,2021A&A...647A..67B},  the ambient medium number density to be $n = 2.8\times 10^{-1} \, \text{cm}^{-3}$ \citep{2007MNRAS.380....2W,2022A&A...664A.166R}, and the jet power to be $P_{\text{jet}} \simeq 1.4\times 10^{44} \ {\rm erg\,s^{-1}}$ \citep{2022A&A...664A.166R}. This yields a stability parameter, $\lambda_{\rm NGC315}\sim2\times10^{-4}-2\times10^{-3}$, one to two orders of magnitude below our models M5 and M5C, as evaluated at $t=20$k. Because the accretion flow in NGC~315 is in the low-luminosity, radiatively-inefficient regime, $L/L_{\rm Edd}\simeq 5\times 10^{-4}\ll1$ \citep{2007ApJ...671L.105G}, our energy-conserving model M5 provides the most appropriate comparison. 
The jet in our model M5 transitions from quasi-parabolic to quasi-conical shape at a factor of a few smaller radius than $r_{\rm B}$. This is very different than what is observed: observations indicate that the parabolic to conical jet transition in NGC~315 happens at sub-pc scales, at $r_{\rm tr}\simeq 5\times 10^3r_{\rm g}\simeq0.6$~pc, which is much smaller than the Bondi scale, $r_{\rm B} \simeq 4\times 10^5r_{\rm g}\simeq50$~pc \citep{2021A&A...647A..67B}. In our models, the near-BH jet collimation is largely controlled by the disk winds, with the jets transitioning from parabolic to conical on the length scale comparable to the radial extent of our thick accretion torus. Thus, the observed parabolic-to-conical transition distance can allow us to infer the radial extent of the inner thick radiatively-inefficient accretion flow (RIAF) in NGC~315: $r_{\rm RIAF}\simeq r_{\rm tr} \simeq 5\times10^3r_{\rm g} \simeq 0.01 r_{\rm B}$. 
Note that similar to M87, at late times, the stability parameter of our M5 model drops down to $\lambda_{\rm M5,60k}\lesssim10^{-3}$. Hence, the late-time evolution of our M5 model might provide a quantitative comparison to NGC~315. We leave such a detailed comparison to future work.

For Cygnus A, we use the BH mass of $M = 2.5 \times 10^9 \, M_\odot$ \citep{2011ApJ...727...39M}, the ambient medium number density of $n = 3 \ \text{cm}^{-3}$ \citep{2016PASJ...68...26F},  the Bondi radius of $r_{\text{B}} = (0.015-0.12) \, \text{kpc}$ \citep{2016PASJ...68...26F,2019ApJ...878...61N},$^\getrefnumber{ftn:rb_def}$ and the jet power of $P_{\text{jet}} = 3.9 \times 10^{45} \, {\rm erg\,s^{-1}}$ \citep{2011ApJ...727...39M}. This results in $\lambda_{\rm CygA}=2\times10^{-4}-0.015$. Since Cygnus A has $L/L_{\rm Edd}\sim 1 \%$ \citep{2015ApJ...808..154R}, disk cooling might be significant: we will compare it with both cooled and energy-conserving runs. At $t =20$k, we have $\lambda_{\rm M5,20k}=0.05$ for model M5 and $\lambda_{\rm M5C} = 0.02$ for model M5C. We conclude that $\lambda$ in the jets in models M5 and M5C is at the upper range of the inferred $\lambda$ range for Cygnus~A.
At late times, $t\simeq 50$k, the jets in model M5C exhibit a similar jet collimation profile to that observed in Cygnus~A: the jets expand quasi-parabolically at small radii and switch to cylindrical inside the Bondi radius \citep{2016A&A...585A..33B}. Interestingly, recent observations indicate that Cygnus~A jet radius seemingly discontinuously increases by an order of magnitude across $r = r_{\rm B}$ and thereafter continues expanding parabolically out to kpc-scales \citep{2019ApJ...878...61N}. The nature of the jet radius jump and kpc-scale parabolic expansion is presently unclear. The last jet shape snapshot shown in deep purple color in Fig.~\ref{fig:rjetm5c} reveals signs of jet recollimation at $r\sim4r_{\rm B}$, i.e., pinching followed by bounce: this feature might be related to the observed discontinuous increase in Cygnus~A jet radius at $r\sim r_{\rm B}$ but does not explain the resumption of parabolic jet shape at $r \gg r_{\rm B}$. Longer-term simulations will be required to investigate the larger-scale jet collimation properties.

\subsection{Limitations and future work}
Whereas our simulations have a high resolution that is enough to resolve the jet radius out to $r = 10^4 \, r_{\text{g}}$ with $5$ cells across the jet radius, we have simplified the setup to make it less numerically expensive. We start with a torus in a hydrostatic equilibrium, which later launches winds that are dynamically important to the jet collimation. However, as \cite{2013A&A...559A.116P} points out, the \cite{1976ApJ...207..962F} prescription of the torus sets the Bernoulli parameter, which determines whether the material is bound and impacts the amount of the disk material that leaves the system via winds. \cite{2013A&A...559A.116P} remark that the final Bernoulli parameter appears to keep the memory of the initial conditions and depends on the initial Bernoulli parameter value prescribed by the torus solution. If a more realistic torus has more tightly bound material, then we expect to see weaker winds and even wider jets inside the Bondi radius. Note that the cooling function decreases the internal energy of the gas and makes it more tightly bound, thereby helping to erase the memory of the initial torus parameters. To investigate how the initial prescription of a torus impacts the power dynamics between disk winds and backflows and, overall, jet evolution, studies of the jet collimation profile for different initial and self-consistently formed tori are needed at the largest scale separations (e.g., \citealt{2021MNRAS.504.6076R}, \citealt{2022ApJ...936L...5L}, \citealt{2023arXiv231011487L}). 

In addition to the accretion disk structure, we assumed a magnetic field configuration, which results in the SANE disk and jets. MAD jets are also possible and may have different collimation profiles: they tend to be more powerful and show more variability, so for the same ambient density and jet power, we expect less collimated jets, less prominent backflows, and stronger time-dependence (\citealt{2003ApJ...592.1042I}, \citealt{2003PASJ...55L..69N}, \citealt{2011MNRAS.418L..79T}, \citealt{2022ApJ...941...30C}).

For our torus in M5C, we use a predefined cooling function \citep{2009ApJ...692..411N} to cool the disk towards the target scale height, $H / R = 0.1$. The cooling function reduces the internal energy of the disk but does not include radiation-driven outflows \citep{2023ApJ...945...57H}, which may strengthen the disk winds and more strongly collimate the jets. It is also important to note that the disk does not reach the target scale height at the outer edge until $t \simeq 60 \text{k}$. In our study, the inner parts of the disk launch the winds and, therefore, contribute significantly to the jet collimation, and the disk in M5C reaches the target scale height by $t = 20 \text{k} \, r_{\text{g}}$ out to $r \simeq 100 \, r_{\text{g}}$, which is the scale of the jet radius. However, the overarching structure of the disk can impact the supply of the magnetic fields and, thereby, impact the power of the jets. As the disk becomes thinner, it could lose its ability to advect large-scale magnetic fields inwards as theorized by \cite{1994MNRAS.267..235L}, which would make the jets weaker. To fully understand the impact of disk cooling on the shape of the jets, 3D radiation-transport GRMHD simulations of sub-Eddington accretion disks are needed \citep[e.g.,][]{2022ApJ...935L...1L,2023ApJ...944L..48L,Liska2024}. 

To be able numerically to follow the jet shape well outside the Bondi radius, we reduce the scale separation and choose $r_{\text{B}} = 10^3 \, r_{\text{g}}$. However, the Bondi radius of a SMBH is of the order of $10^5-10^6 \, r_{\text{g}}$ (e.g., \citealt{2015MNRAS.451..588R}). This assumption implies that the ambient medium in our simulations has a higher internal energy, making it harder for jets to propagate. The opening angle of the jets at $r_{\text{B}}$ is set by the collimation of the jets by disk winds. For more realistic values of the Bondi radius, the winds will have a larger range of distances over which to collimate the jets: the jets will become more collimated and, potentially, faster. If numerically possible, simulations with larger values of the Bondi radius will shed light on the scale separation dependence of the problem.

In our simulations, we adopt constant ambient density and pressure profiles to evaluate at the zeroth order the impact of the ambient medium on the jet collimation. However, the radial ambient density profiles in AGN typically follow a power-law: for instance, in M87, the density around the Bondi radius scales as $\rho \propto r^{-1}$ \citep{2015MNRAS.451..588R}. Such a density profile is effectively equivalent to a less dense ambient medium, resulting in jets turning into cylinders at larger $r$; we expect the jets to propagate faster and be less collimated. Depending on the radial power-law index of the ambient medium distribution, the backflows might not form at all since, at the jet head, the ambient medium can be negligible compared to the jet, unable to redirect the jet material, and might not be able to turn the jets into cylinders. Moreover, we expect that in real systems, the density profile is highly inhomogeneous due to the previous jet activity and external factors \citep{2012ARA&A..50..455F}. As the ambient medium plays a key role in the jet collimation, in future work, we will explore how different ambient medium density profiles impact the shape of the jet.

In Section \ref{subsec:rjet}, we measure the jet shapes in models M7, M6, and M5 at $t \simeq 20$k. The jet shapes show little variability at this time. This makes our measurements robust. However, we expect that, as time progresses, the radii of transitions between the parabolic and the conical regions, as well as between the conical and cylindrical regions, will move outward as Figure \ref{fig:rjetm5c} shows for model M5C. However, in model M5, the jet becomes energetically dominated by the disk winds, which sets a characteristic timescale of jet evolution. In future work, we will study the time dependence of jet shape break radii to understand how it maps onto the observations of jets whose age is known (e.g., 3C~84 jet).

\section{Summary}
\label{section-summary}

In this Letter, we present 3D GRMHD simulations of BH-launched jets interacting with disk winds and constant density ambient medium, at the largest Bondi-to-BH scale separation to date, $r_{\text{B}} = 10^3 r_{\text{g}}$. We find that when we increase the ambient medium density by $2$ orders of magnitude, the jets exhibit parabolic shapes and become conical at $r \simeq 200 \, r_{\text{g}}$, independent of the ambient medium density: this is because the radial extent of the accretion disk determines the near-BH collimation profile of the jets. Formed due to the jet interaction with the ambient medium, backflows create a constant-pressure jet environment at larger distances, $r \gtrsim r_{\text{B}}$, e.g., $r \simeq 6r_{\text{B}}$ for model M7, and turn the jets into cylinders. The conical-to-cylindrical transition point moves inward for higher ambient medium densities, or for less powerful jets relative to the environment: they struggle to drill through the ambient medium and form more prominent backflows that collimate the jets into cylinders at smaller radii. For instance, when the jet to ambient energy ratio is lower by a factor of $10$ and $100$, the jets turn into cylinders at $r \simeq 3r_{\text{B}}$ and $r \simeq r_{\text{B}}$, respectively.

We considered AGN, such as M87 and NGC~315, that have radiatively inefficient accretion flows. We found that although our total energy-conserving models tended to feature more powerful jets than in these systems, they allowed us to interpret the observations qualitatively. We speculate that the jet in M87 might fall apart due to the kink instability before forming backflows, and therefore, it does not have the cylindrical collimation we have seen in models M7, M6, and M5. Our simulations results suggest that the parabolic to conical jet shape transition occurs where the thick disk ends. Thus, the surprisingly small distance the jet shape transition in NGC 315 (much smaller than $r_{\rm B}$) might be related to the size of the thick accretion disk in this system.

However, none of these total energy-conserving (nonradiative) models results in cylindrical jets well inside $r_{\text{B}}$, unlike what is seen in Cygnus A and 3C 84. Because Cygnus A and 3C 84 accretion flows are expected to be radiatively efficient, we explored if disk cooling can affect the jet collimation profile by prescribing a cooling function in our model M5C. Radiative cooling saps energy from the disk winds and weakens them. This enables the jet in our model M5C to energetically dominate the disk winds, conically expand in a ballistic fashion, and become exceptionally wide close to the BH. The weaker disk outflows allow jet backflows, formed by the jet-ambient-medium interaction, to reach deep inside the Bondi radius and collimate the jets into cylinders already at $r \simeq 300 \, r_{\text{g}}$, similar to the 3C~84 jet. Over time, the disk winds become more powerful and cause the jet to become quasi-parabolic near the base and cylindrical father out but still inside the Bondi radius, similar to Cygnus A. Summing up, whereas our model M5C is simplified, it offers a controlled experiment that reveals a qualitatively new, long-sought behavior: conical jet expansion near the BH and backflows collimating jets into cylinders well inside $r_{\text{B}}$.

\section*{Acknowledgments}
VR and MP were supported by a grant from the Baker Program in Undergraduate Research.
MP acknowledges support from the Illinois Space Grant.
KC is supported in part by grants from the Gordon and Betty Moore Foundation and the John Templeton Foundation to the Black Hole Initiative at Harvard University, and by NSF award OISE-1743747.
ML was supported by the John Harvard, ITC, and NASA Hubble Fellowship Program fellowships.
AT acknowledges support from the NSF AST-2009884 and NASA 80NSSC21K1746 grants.
AT was supported by BSF grant 2020747 and NSF grants AST-2107839, AST-1815304, AST-1911080, AST-2206471, OAC-2031997.
OG is supported by Flatiron Research and CIERA Fellowships.
OG also acknowledges support by Fermi Cycle 14 Guest Investigator program 80NSSC22K0031.
The authors acknowledge the Texas Advanced Computing Center (TACC) at
The University of Texas at Austin for providing HPC and visualization
resources that have contributed to the research results reported
within this paper via the LRAC allocation AST20011
(\url{http://www.tacc.utexas.edu}).
This research used resources from the Oak Ridge Leadership Computing Facility, which is a DOE Office of Science User Facility supported under Contract DE-AC05-00OR22725. An award of computer time was provided by the ASCR Leadership Computing Challenge (ALCC), Innovative and Novel Computational Impact on Theory and Experiment (INCITE), and OLCF Director's Discretionary Allocation programs under award PHY129.
This research used resources of the National Energy Research Scientific Computing Center, a DOE Office of Science User Facility supported by the Office of Science of the U.S. Department of Energy under Contract No. DE-AC02-05CH11231 using NERSC awards ALCC-ERCAP0022634 and m2401 for 2022 and 2023. It also used the computational resources of Calcul Quebec (http://www.calculquebec.ca) and Compute Canada (http://www.computecanada.ca).




\end{document}